# Purcell-enhanced single-photon generation from $CsPbBr_3$ quantum dots in in-situ selected Laguerre-Gaussian modes

Virginia Oddi[1,2], Darius Urbonas[1], Etsuki Kobiyama[1], Ioannis Georgakilas[1], Ihor Cherniukh[2,3], Kseniia Shcherbak[2,3], Chenglian Zhu[2,3], Maryna I. Bodnarchuk[2,3], Maksym V. Kovalenko[2,3], Rainer F. Mahrt[1], Gabriele Rainò[2,3], Thilo Stöferle[1]

[1] IBM Research Europe – Säumerstrasse 4, 8803 Rüschlikon, Switzerland

[2] Department of Chemistry and Applied Biosciences, ETH Zurich – Vladimir Prelog Weg 1, 8093 Zürich, Switzerland

[3] Laboratory for Thin Films and Photovoltaics, Empa, 8600 Dübendorf, Switzerland

## ABSTRACT

Single photons in Laguerre-Gaussian (LG) beams, which carry orbital angular momentum (OAM), could enable more robust and efficient photonic quantum communication and information processing, as well as enhanced sensitivity in quantum metrology and imaging. However, as most implementations are indirect or require additional mode-shaping elements, direct generation of single photons with OAM has received growing interest. Colloidal lead halide perovskite quantum dots (QDs) have recently emerged as a versatile material that can produce indistinguishable single photons quasi-deterministically at high rate. Here, we integrate single $CsPbBr_3$ QDs into an open Fabry-Pérot microcavity with a nanofabricated Gaussian-shaped deformation, demonstrating Purcell-enhanced single-photon generation into individual cavity modes with up to $18.1 \pm 0.2$ times accelerated decay, down to tens of picoseconds. By in-situ tuning of the cavity resonance, we can selectively couple a single QD to different LG modes carrying OAM and observe the spatial patterns of the generated single-photon beams emitted from the cavity. Our findings open the door to high photon rate sources that directly generate single-photon LG beams for advanced quantum photonic applications.

**MAIN TEXT**

The generation of single photons lies at the heart of quantum technologies. From ultra-secure communication to quantum computing and precision sensing, on demand single-photon sources are essential building blocks. Around the world, researchers are in a race to develop reliable, on-demand single-photon emitters, pushing the boundaries of material science, nanotechnology, and photonic engineering to meet the growing demands of scalable quantum systems[1–4]. To increase the extraction efficiency and photon quality, single quantum emitters are typically integrated into optical microcavities. The cavity modifies the density of photonic states and thereby, through the Purcell effect, boosts the spontaneous emission rate as described by Fermi's golden rule. The emission can be coerced into a single, well-defined cavity mode, effectively enhancing brightness, purity and, due to the accelerated decay, also coherence and indistinguishability of the emitted photons[5,6]. When the emitter is both spectrally and spatially resonant with the cavity mode[7,8], the enhancement in emission rate is maximized and quantified by the Purcell factor, $F_P = 3Q\lambda^3/4\pi^2 n^3 V$, where $Q$ is the quality factor, $V$ the effective modal volume, $n$ the refractive index and $\lambda$ the resonance wavelength of the cavity

Epitaxially grown III-V semiconductor quantum dots (QDs) represent the most exploited solid-state system for quasi-deterministic single photon generation. Successful implementation using epitaxially-grown QDs embedded in microcavities exhibit a large Purcell factor[9–11] of up to several tens and accomplish a high purity, coherence and indistinguishability[6,12–14]. In a major step towards ideal deterministic sources, tunable, open Fabry-Pérot microcavities with a micrometer-sized hemispherical deformation in one of the mirrors, allowing in-situ selection of the best QDs, have recently been used to achieve GHz rate generation of high-quality single photons[15] and system efficiency exceeding 70%[16].

Colloidal lead halide perovskite QDs with easy and cost-effective solution processability through wet-chemical synthesis[17], have attracted increasing attention as single-photon sources that could potentially play a pivotal role in the development of next-generation photonic quantum technologies[18,19]. In the strong electronic confinement regime, perovskite QDs have achieved single-photon emission purity as high as 98% without spectral filtering[20]. Weakly confined $CsPbBr_3$ QDs at cryogenic temperature, owing to their large oscillator strength, exhibit an extraordinarily fast radiative decay, reaching sub-100 ps lifetime[21], about one order of magnitude faster than self-assembled III-V semiconductor QDs. Moreover, emission from single $CsPbBr_3$ QDs has shown coherence times of 80 ps (with a radiative lifetime of 210 ps)[22] and a photon indistinguishability of 56%[23]. So far, however, only few studies have explored the integration of perovskite QDs into optical microcavities, mostly on ensemble level[24–27]. Single QDs in a tunable, open microcavity have shown reduced emission linewidth at room temperature[28]. However, even at cryogenic temperatures, Purcell factors remained below 2 in circular Bragg grating resonators[29] and tunable fiber-cavities[30].

Numerous advanced quantum photonic applications exploit the orbital angular momentum (OAM) of photons, such as realizing high-dimensional entanglement for more efficient and robust quantum communication, encoding qudits[31] or precision measurements[32]. Single photons from spontaneous parametric down-conversion or epitaxial QDs are typically transformed into freely adjustable OAM modes, such as Laguerre-Gaussian (LG) beams, employing mode-shaping elements. However, this reduces efficiency and hinders integration. Approaches for direct generation of single photons with OAM using colour centres in lithographically defined metasurfaces[33,34] and epitaxial QDs in ring resonators[35] have not been able to benefit from the high Purcell enhancement in wavelength-scale microcavities with high $Q/V$ and do not allow in-situ mode selection.

In this work, we integrate single $CsPbBr_3$ QDs into a tunable, open Fabry-Pérot microcavity with dielectric mirrors comprising a nanofabricated Gaussian-shaped deformation. By comparing the time-resolved emission decay of the same QDs inside and outside of the cavity, we demonstrate a Purcell enhancement of up to $18.1 \pm 0.2$ at 50 K, accompanied by increased brightness. By using in-situ length-tuning of the cavity to bring LG modes of different radial and azimuthal order into resonance with the QD emission, we can coerce the QD to generate single photons directly into these selected LG modes, as evidenced by imaging of the distinct modal patterns.

In our experiments, we use chemically synthesized colloidal $CsPbBr_3$ nanocrystals (see Methods) with a size of $25.3 \pm 1.5$ nm, determined from statistical image analysis of high-resolution transmission electron microscope (HRTEM) images as the one shown in Figure 1a. When dispersed in toluene, these QDs exhibit a photoluminescence (PL) quantum yield (QY) of 65% at room temperature. The room-temperature PL emission spectrum, shown in Extended Data Fig. 1a, alongside the corresponding absorption spectrum, features a peak energy of 2.39 eV and a full width at half maximum (FWHM) of 76 meV. When spin-coated onto a substrate and cooled to cryogenic temperatures, the emission red-shifts, and the linewidth drastically narrows down to sub-meV at 6 K for individual emitters. The fine structure of the bright triplet excitons in single QDs can be clearly resolved, with up to three distinct peaks (Fig. 1b) of dominantly linear polarization (inset), depending on the crystal orientation and observation direction[36,37]. Single-photon emission is established by the characteristic anti-bunching observed in the second-order correlation function at zero-time delay, $t$, $g^{(2)}(t=0)$, as presented in Fig. 1c, where the raw, uncorrected data (brown line) are fitted with a double-sided exponential function (blue line). The measured value of $g^{(2)}(0) = 0.13 \pm 0.02$ confirms single-photon emission, while the relatively high value arises from residual contributions of (charged) multi-exciton emission.

The single QDs are placed into a tunable Fabry-Pérot-type open microcavity (see Methods) with a Gaussian-shaped deformation[38] to provide tight lateral confinement of the mode, as schematically shown in Fig. 1d. The use of two separate, independent mirrors allows not only for length tuning of the cavity but also for easy repositioning of the Gaussian deformation over different QDs as well as for removing the top cavity half to study the same QDs without microcavity for direct comparison. Due to the cylindrical symmetry, this cavity configuration supports Laguerre-Gaussian modes $LG_{nl}$, denoted by their radial, $n$, and orbital, $l$, angular momentum quantum numbers[39,40]. Fig. 1e shows a three-dimensional finite-difference time-domain simulation (3D FDTD) of the energetically lowest mode, $LG_{00}$, from which we obtain a quality factor of $Q = 1941$ with a resonant wavelength of $\lambda = 533$ nm and an effective modal volume of $V = 3.3\ \lambda^3$ with an air gap of 705 nm in the cavity centre. This configuration results in a calculated Purcell factor of 37.9, see Methods. To investigate the optical properties of the fabricated cavity and the in-situ tunability of resonant modes by changing the distance between the cavity halves, we employ spectroscopic white-light reflection measurements (Fig. 1f), from which we obtain a $Q$-factor of 623, corresponding to a 3.75 meV FWHM at a peak energy of 2.338 eV (inset). We operate the cavity in a mode where the two halves are slightly tilted from perfect parallelism and get in contact far outside of the central region with the Gaussian deformation, drastically reducing the mechanical vibrations and cavity length jitter while maintaining tunability[41]. Yet, the considerably lower apparent *Q*-factor compared to the theoretical calculation may be due, at least in part, to broadening caused by residual vibrations. As the air gap cannot be measured directly, we perform transfer-matrix simulations of a planar cavity that allow to correlate the observed cavity resonance tuning to the actual cavity length. To account for the Gaussian deformation, which effectively increases the local cavity length, the calculated planar modes are shifted (white points for $LG_{00}$ and red points for $LG_{01}$), as detailed in Extended Data Fig. 1c.

**Purcell-enhanced single-photon emission**

To investigate cavity-enhanced single photon emission in $CsPbBr_3$ QDs, we perform measurements at temperatures of 6 K (Fig. 2a-c) and 50 K (Fig. 2d-f). We show the radiative decay (Fig. 2a) and PL spectrum (Fig. 2b) of an exemplary QD (QD#1) positioned inside (red line) and outside (grey line) the cavity at 6 K. The instrument response function (IRF) curve of the single-photon detector is also included in Fig. 2a and displayed as blue line. When the QD emission is resonant with the cavity, the emission is guided into a specific LG mode (inset of Fig. 2b), and an acceleration in the radiative decay, along with an enhancement in the emitted intensity is observed. Extracted lifetimes of three QDs (QD#1, 2, 3) are presented in Fig. 2c, with a maximum Purcell factor $F_P = \tau_{fast,out} / \tau_{fast,in} = 4.2 \pm 0.1$, quantified by fitting the decay traces with a double-exponential function, where $\tau_{fast,in/out}$ denotes the time constant of the fast component in- and outside of the cavity. The slow component, $\tau_{slow}$, is not Purcell-accelerated as it is attributed to delayed refilling from trap states[42] or, in some instances, limited by the long tail of the IRF. As shown by the comparison with the IRF trace, a potentially much faster intrinsic acceleration cannot be resolved from this data, as the cavity-accelerated decays at 6 K approach the temporal resolution of the detector.

This experimental limitation can be circumvented when measuring those QDs at intermediate temperature, as the radiative decay time in perovskite QDs increases with increasing temperature due to reduction of the single-photon superradiance effect[21], while the emission linewidth remains sufficiently narrow to couple well to the cavity mode. Therefore, we present measurements for another exemplary QD (QD#4) at 50 K, showing the radiative decay (Fig. 2d) and PL spectrum (Fig. 2e), both inside and outside of the cavity, along with the corresponding real space mode profile (inset of Fig. 2e). The relative speed-up in the decay trace is much more pronounced compared to lower temperature, but the fast component still remains close to the IRF limit. Additionally, compared to 6 K, the stronger enhancement in the emitted intensity at 50 K may be due to a change in the effective QY, as the Purcell effect can accelerate the radiative decay within the time scale of non-radiative quenching processes. The extracted lifetimes for three QDs (QD#4, 5, 6) are presented in Fig. 2f, showing a maximum $F_P$ = 18.1 ± 0.2. Inside the cavity, the lifetime values are again obtained from a double-exponential fit, while outside of the cavity a single-exponential fit is sufficient, as the QD decay becomes similar or longer than the slower time scale. The emission decay, spectrum and real space mode profiles of the additional QDs included in the charts of Fig. 2c and Fig. 2f are displayed in Extended Data Fig. 2 and Extended Data Fig. 3, respectively. Additionally, an overview of the extracted parameters $\tau_{fast}$, $\tau_{fast,out}$ / $\tau_{fast,in}$, $\tau_{slow}$ and the ratio of the spectrally integrated emission intensity inside and outside the cavity, $A_{in}$ / $A_{out}$, for QD#1-6 is provided in Extended Data Table 1.

The maximum measured Purcell enhancement is lower than the calculated value of 37.9. This may be attributed to imperfections in the cavity fabrication, slightly suboptimal spectral or spatial alignment of the emitter with the optical mode or to the quality of QD emission, as finite linewidth and a low temporal stability and/or spectral diffusion would reduce the coupling between the QD and the cavity mode. The PL time-series of two presented QDs in Extended Data Fig. 4c and Extended Data Fig. 5 show energy drift over timescales of a few seconds and fluctuations in the emitted intensity.

An additional benefit of Purcell-enhanced coupling of an emitter to a single optical cavity mode stems from the intrinsically provided spectral filtering, eliminating the need for (lossy) external filters. This is particularly relevant for $CsPbBr_3$ QDs where spectral filtering is important to suppress biexciton emission, which significantly reduces the photon purity[21,22] due to its high QY in large QDs at low temperature. This is observed in our results as plotted in Extended Data Fig. 4a, b, where we present the $g^{(2)}(0)$ of QD#1 outside and inside the cavity. Outside the cavity, when a tunable bandpass filter is used to suppress the biexciton emission, we observe a $g^{(2)}(0) = 0.25$ (pink line) and close to 1 when the filter is removed (grey line). However, in the presence of the cavity, $g^{(2)}(0)$ reaches 0.4 even without the use of an additional filter.

**Generation and control of photons in LG modes with OAM**

Outside of the cavity, the QD emission is observed as a diffraction-limited Gaussian spot, as seen in Fig. 3a. In the experiments above, we had tuned the $LG_{00}$ mode into resonance with the QD, also resulting in a Gaussian-shaped cavity emission. However, the cylindrically symmetric cavity also supports higher-order LG modes with non-zero OAM, see schematic in Fig. 3b. Hence, when placing a QD inside the cavity and length-tuning the cavity to bring such modes into resonance with the QD, the single photon emission will occur directly into these LG modes. Due to the linearly polarized fine structure lines that make up the excitonic emission of perovskite QDs, both positive and negative OAM quantum numbers can be excited simultaneously; for example, the photons will be in a superposition of left- and right-helical modes. The orientation of the linear dipole is reflected by the phase difference of both states that results in an observed dipole-like pattern in case of $LG_{01}$ instead of the donut mode shape, as illustrated in Fig. 3c. Dependent on which fine structure line is chosen to overlap best with the cavity resonance, the orientation of the dipole pattern can change because of their different linear polarization.

The measured emission profiles of different $LG_{nl}$ modes resulting from the coupling with single QDs at different cavity lengths are shown in Fig. 3 (d, e, top row) and compared with the calculated corresponding eigenstates (d, e, bottom row), see Methods for details. In Fig. 3d, $LG_{00}$, $LG_{01(90°)}$, $LG_{01(0°)}$ and $LG_{02(90°)}$ are extracted from the same single QD, demonstrating the in-situ tunability of our system that enables the selection of a specific LG mode and therefore OAM superposition state. In Fig. 3e, $LG_{11(90°)}$ and $LG_{11(45°)}$ originate from the same QD (but different QD than for Fig. 3d), whereas $LG_{03(0°)}$ and $LG_{20}$ are obtained from different QDs. Due to the random orientation of the individual QDs and therefore of the fine structure polarization, some QDs couple more efficiently to certain LG modes than others, and slight lateral spatial displacement of the QDs can also change the coupling efficiency. The spectrum of each $LG_{nl}$ mode is shown in Extended Data Fig. 6.

**Conclusions**

In summary, we employed a tunable, open microcavity with an engineered Gaussian-shaped deformation to couple single colloidal $CsPbBr_3$ QDs to well-defined $LG_{nl}$ modes at 6 K and 50 K. We observed Purcell-accelerated emission decays on the order of 30 ps, limited by the temporal resolution of our detector, and a measured Purcell factor of up to $18.1 \pm 0.2$ at 50 K. We demonstrate the direct generation of single photons into LG modes carrying OAM that can be controlled by in-situ tuning of the cavity resonance. These findings lay the groundwork for a viable strategy for high-efficiency single-photon sources for quantum applications that utilize the additional dimensions provided by LG states.

**Methods**

**QD synthesis and basic characterization.** The synthesis of the 25-nm CsPbBr$_3$ dots was achieved through the modification of the TOPO-PbBr$_2$ approach[43]. The synthesis was performed by slowly injecting Cs and PbBr$_2$ stock solutions into mesitylene. After injection, the solution was treated with a zwitterionic ligand, 2-ammonioethyl (hydroxypolypropylene glycolyl) phosphate (PPG-PEA), which was synthesized according to the procedure described by Morad *et al.*[44]. The resulting solution was then precipitated with hexane and redispersed in toluene. The synthesis details will be published elsewhere. The HRTEM image shown in Fig. 1a was collected using a JEOL JEM-2200FS microscope operated at 200 kV.

For the further sample preparation, CsPbBr$_3$ QDs dispersed in solution are diluted up to a factor $10^4$ in toluene starting from a concentration of ~1 mg/ml. For basic QD characterization shown in Fig. 1b, c, the diluted solution is then spin-coated onto a 10x10 mm crystalline Si wafer covered with a 3-μm-thick thermal-oxide layer. All samples are prepared in gloveboxes under argon or nitrogen atmosphere.

**Microcavity fabrication.** The dielectric Fabry-Pérot microcavity consists of two independent cavity halves. The top part builds on a glass substrate with a mesa, designed to minimize the contact area and sensitivity to particle contamination, allowing to reach gaps between the cavity halves of few hundred nanometers. The mesa structure is fabricated using optical lithography followed by wet etching with concentrated HF, resulting in a mesa that is approximately 30 μm tall and 250 μm wide. A Gaussian-shaped deformation of ~ 60 nm depth and ~ 4 μm FWHM (Extended Data Fig. 1b) is patterned on top of the mesa using Focused Ion Beam (FIB) milling. Subsequently, 7.5 distributed Bragg reflector (DBR) layer pairs are deposited conformally via Ion Beam Deposition (IBD), composed of alternating quarter-wave layers of SiO$_2$ and Ta$_2$O$_5$. For the bottom cavity half, 9.5 DBR pairs are deposited on a 20 mm × 20 mm Si substrate via IBD followed by an 85 nm-thick SiO$_2$ spacer layer using an e-beam evaporator onto which the diluted QD solution is spin-coated.

**Optical characterization.** The two halves of the microcavity are mounted inside a cold-finger liquid-helium-flow cryostat on separate *xyz* nano-positioning stages, with additional tilt stages for the upper half. A home-built µ-PL setup is used for spectroscopic measurements of single QDs and the microcavities. A mode-locked Ti:Sa oscillator (Tsunami, Spectra Physics) with a repetition rate of 80 MHz and pulse duration of 100 fs serves as excitation source after frequency-doubling to 3.06 eV photon energy through a barium borate (BBO) crystal. The light is then guided to the µ-PL setup via an optical single-mode fiber, stretching the pulses in time to several picoseconds. The excitation power is adjusted after the fiber with a graduated neutral-density filter mounted on a motorized linear stage and monitored after a beam splitter (BS) pick-off by a power meter. After passing through a dichroic BS (463 nm edge wavelength, Semrock), the excitation beam is focused on the sample through a microscope objective (Mitutoyo 100X apochromat, NA=0.7), reaching a 1/e$^2$ diameter of 1.9 µm. Typical fluences used to excite single QDs are 0.1-0.85 µJ cm$^{-2}$. The PL emission is collected from the same objective lens and passes through the same dichroic BS and a long-pass filter (450 nm edge wavelength, Semrock). Additionally, a tunable band-pass filter (Semrock) can be inserted as needed. Then, a flip mirror directs the light either to a Quantitative-CMOS camera (ORCA-Quest, Hamamatsu Photonics) or to a 750 mm-long monochromator (Acton) equipped with a grating with 1800 lines/mm and an EMCCD detector (ProEM, Princeton Instruments). For $g^{(2)}$ and lifetime measurements, either a 50/50 BS (for simultaneous acquisition with spectra) or a mirror is placed in front of the monochromator to send the light to a Hanbury Brown-Twiss setup with an unpolarized 50/50 BS and two avalanche photodiodes (PDM, Micro Photon Devices) that are recorded with a time-correlated single-photon counting system (PicoHarp, PicoQuant). For the basic empty cavity reflection characterization in the same setup at room temperature, a fiber-coupled halogen lamp is used as excitation source, which is focused on the sample to a spot diameter of 5.1 µm.

To achieve optimal coupling between the QD and the cavity mode, a systematic alignment procedure is followed. First, the Gaussian deformation is precisely positioned over the QD and brought into focus. Once aligned, the cavity length is gradually reduced by moving the bottom part up until the QD and cavity mode are spectrally in resonance and the emitted intensity in the spectrum is maximized. After the measurements with the QD in the cavity, the top part is fully removed from the beam path, allowing the study of the same QD outside the cavity. As can be seen in the work of *Zhu et al.*[21], the decay time for single $CsPbBr_3$ QDs of the same size at low temperature can vary significantly, with differences exceeding 100 ps and reaching up to a two-fold variation at 6 K. Hence, the possibility to make this direct comparison of the same QDs in- and outside of the cavity is important to obtain reliable and precise Purcell factors.

**Photonic simulations.** For the numerical 3D FDTD simulations, we use the commercial software package Lumerical with the refractive indices of the materials experimentally obtained from variable-angle spectroscopic ellipsometry (Wollam VASE). To calculate the effective modal volume $V$, we use the expression $V = \int_V \varepsilon(r)|E(r)|^2 d^3r / \max(\varepsilon(r)|E(r)|^2)$, where $E(\mathbf{r})$ is the electric field and $\varepsilon(r)$ is the electric permittivity. The theoretical Purcell factor is calculated using $F_p = 6\pi c^3 Q / \omega_c^3 V$, where $c$ is the speed of light, $\omega_c$ is the angular frequency of the optical mode and $Q$ the quality factor. This definition considers the optimal case in which the dipole is spectrally narrow and resonant with the cavity mode, located at the electric field antinode and aligned with the local electric field.

We use transfer matrix simulations to calculate the reflectance of the multilayer structure as a function of the cavity length, defining a central wavelength of 530 nm for the DBR stopbands. The model assumes normal incidence and accounts for the wavelength-dependent dispersion of the refractive indices of $Ta_2O_5$ and $SiO_2$. The cavity length is defined as the thickness of the air gap between the flat DBR mirrors.

In order to calculate the LG modes in the Gaussian deformation, we solve the 2D time-independent Schrödinger equation for the photon as particle in the Gaussian potential well, using the QMsolve Python package. We assume an effective mass of $m_{eff} = E / c^2$, with $E$ being the photon energy of 2.32 eV and *c* the speed of light. The confining potential is defined as a Gaussian well with a depth of 94 meV (corresponding to 60 nm actual depth at ~700 nm cavity length, see caption of Extended Data Fig. 1) and a FWHM of 4 μm. To implicitly slightly lift the degeneracy of the cylindrically symmetric eigenstates, we introduce a small anisotropy, up to 2%, by modifying the Gaussian's ellipticity in the calculation. This is used to effectively account for the linear QD dipole orientation, which when coupled to the LG cavity modes leads to photon emission into superposition states of left- and right-handed helicity whose phase relation reflects the azimuthal orientation of the linearly polarized dipole.

**Data availability**

Data supporting the findings of this study are available from the corresponding authors upon reasonable request.

**Acknowledgements**

We thank the team of the IBM Binnig and Rohrer Nanotechnology Center for support with the cavity fabrication. We acknowledge funding from EU H2020 EIC Pathfinder Open project "PoLLoC" (Grant No. 899141), EU H2020 EIC Pathfinder Open project "TOPOLIGHT" (Grant No. 964770), EU H2020 MSCA-ITN project "AppQInfo" (Grant No. 956071), EU H2020 MSCA-ITN project "PERSEPHONe" (Grant No. 956270), Swiss National Science Foundation projects "Q-Light – Engineered Quantum Light Sources with Nanocrystal Assemblies" (Grant No. 200021_192308) and "Cavity-Enhanced Many-Body Interactions in Deterministically Positioned Perovskite Quantum Dots (CHIP-QD)" (Grant No. 200021-231778), and by the Air Force Office of Scientific Research under award number FA8655-24-1-7064.

**Author contributions**

V.O performed the experiments and analysed the data, supported by E.K., I.G. and D.U.. D.U. and I.G. fabricated the cavity structure. I.C. and K.S. synthesized the CsPbBr$_3$ QDs, M.I.B. performed TEM characterization, and C.Z. did the basic optical characterization of the single QDs. T.S. and V.O. performed the simulations. G.R., M.I.B., M.V.K., R.F.M., and T.S. supervised the work and acquired the funding. V.O, T.S and R.F.M. wrote the manuscript with contributions from all the authors.

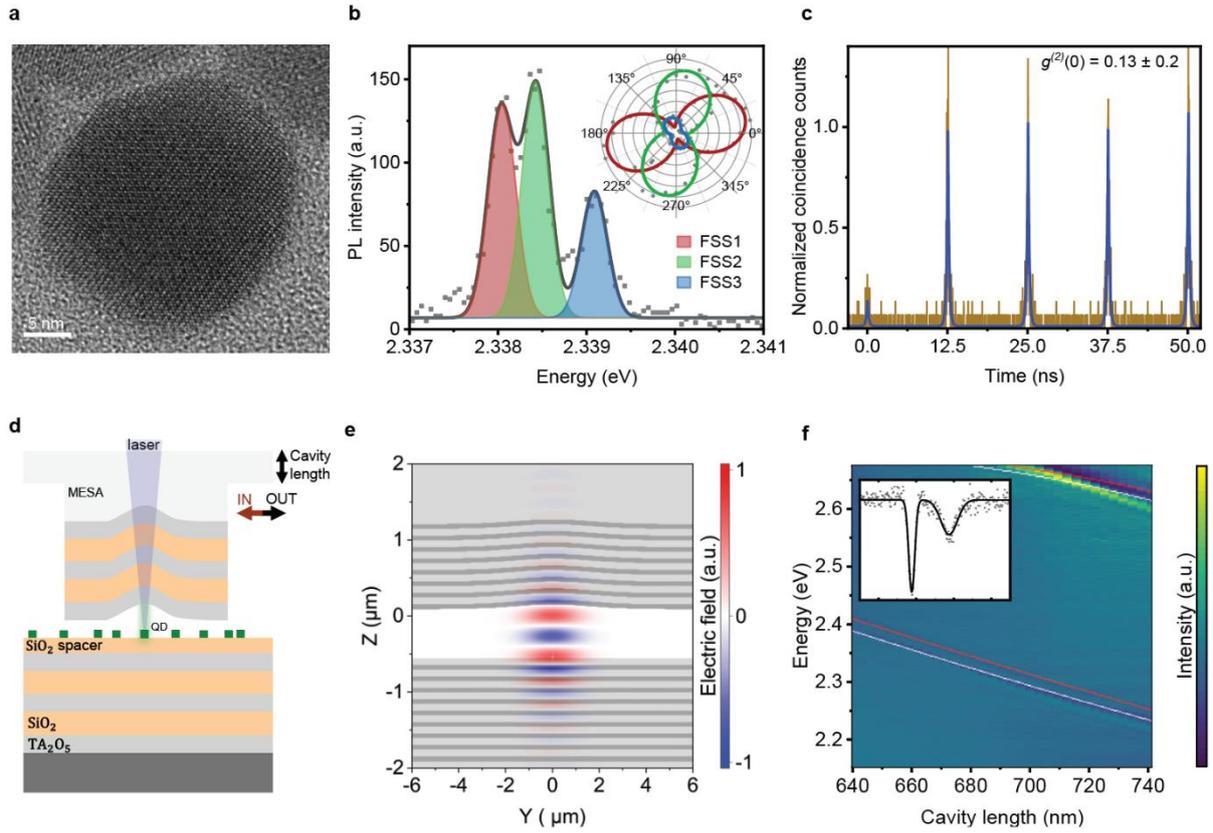

**Fig. 1 | QD emission properties and cavity system. a,** High-resolution transmission electron microscope image of a single 25 nm $CsPbBr_3$ QD. **b,** PL spectrum of a single $CsPbBr_3$ QD at 6 K, fitted with a three-Gaussian-peak function (black solid line) from which the intensity of each fine structure state can be extracted. The intensity of each fine structure line at different angles of a linear polarizer is shown in the inset where the solid lines represent a $\sin^2$ fit. **c,** $g^{(2)}(0)$ measurement, showing a value of $13 \pm 2$ %. It is retrieved from a double-sided exponential fit (blue line) to the raw, uncorrected data (brown line). The exciton emission is filtered with a 15 nm-wide tunable bandpass filter. **d,** Schematic of the tunable open Fabry-Perot cavity system. **e,** Electric field distribution of the $LG_{00}$ mode obtained with 3D FDTD simulation. **f,** Spectroscopic white-light reflection measurement of the bare cavity, showing the tuning of LG mode resonances while changing the cavity length. The cavity length is retrieved from transfer-matrix simulations with the two planar modes of different longitudinal order shifted horizontally to account for the Gaussian potential (white points for $LG_{00}$ and red points for $LG_{01}$), see Extended Data Fig. 1c for more details. A maximum Q-factor of 623, corresponding to a 3.75 meV FWHM at a peak energy of 2.338 eV, is extracted from a Gaussian fit (inset).

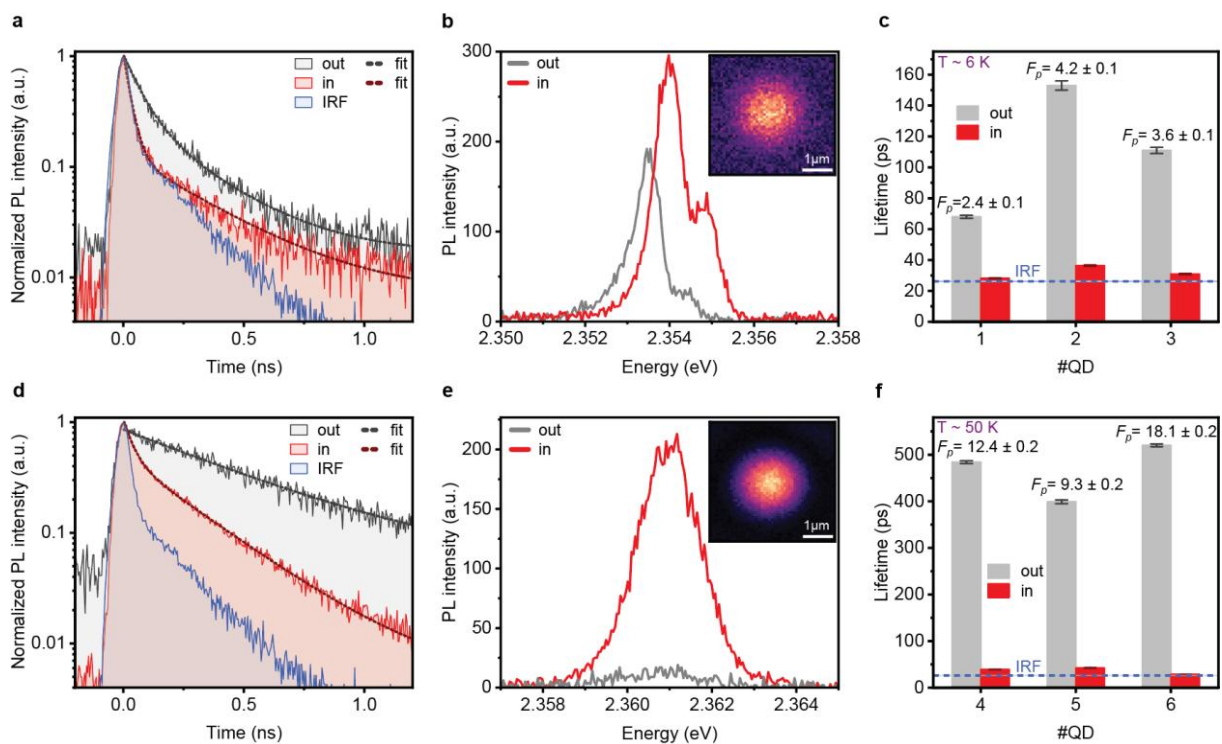

**Fig. 2 | Purcell enhancement at 6 K and 50 K. a, b,** Measurement results at 6 K. Radiative decay **(a)** and PL spectrum **(b)** of a QD (#1 in **(c)**) placed inside (in, red line) and outside (out, grey line) the cavity. The inset in **(b)** shows the real-space $LG_{00}$ mode profile. **c,** Lifetimes and corresponding Purcell factors, $F_P$, of three QDs at 6 K, extracted from the fast component of a double-exponential fit. The indicated instrument response function (IRF, blue dash line) highlights that the Purcell-enhanced lifetimes are close to the temporal detection limit. A maximum Purcell enhancement of $4.2 \pm 0.1$ is reached at 6 K. **d, e,** Measurement results at 50 K. Radiative decay **(d)** and PL spectrum **(e)** of a QD (#4 in **(f)**) placed inside (in, red line) and outside (out, grey line) the cavity. The inset in **(e)** shows the real-space $LG_{00}$ mode profile. **f,** Lifetimes and corresponding Purcell factors, $F_p$, of three QDs at 50 K. Inside the cavity, values are obtained from the fast component of a double-exponential fit, while, outside, from a single-exponential fit. A maximum Purcell enhancement of $18.1 \pm 0.2$ is reached at 50 K.

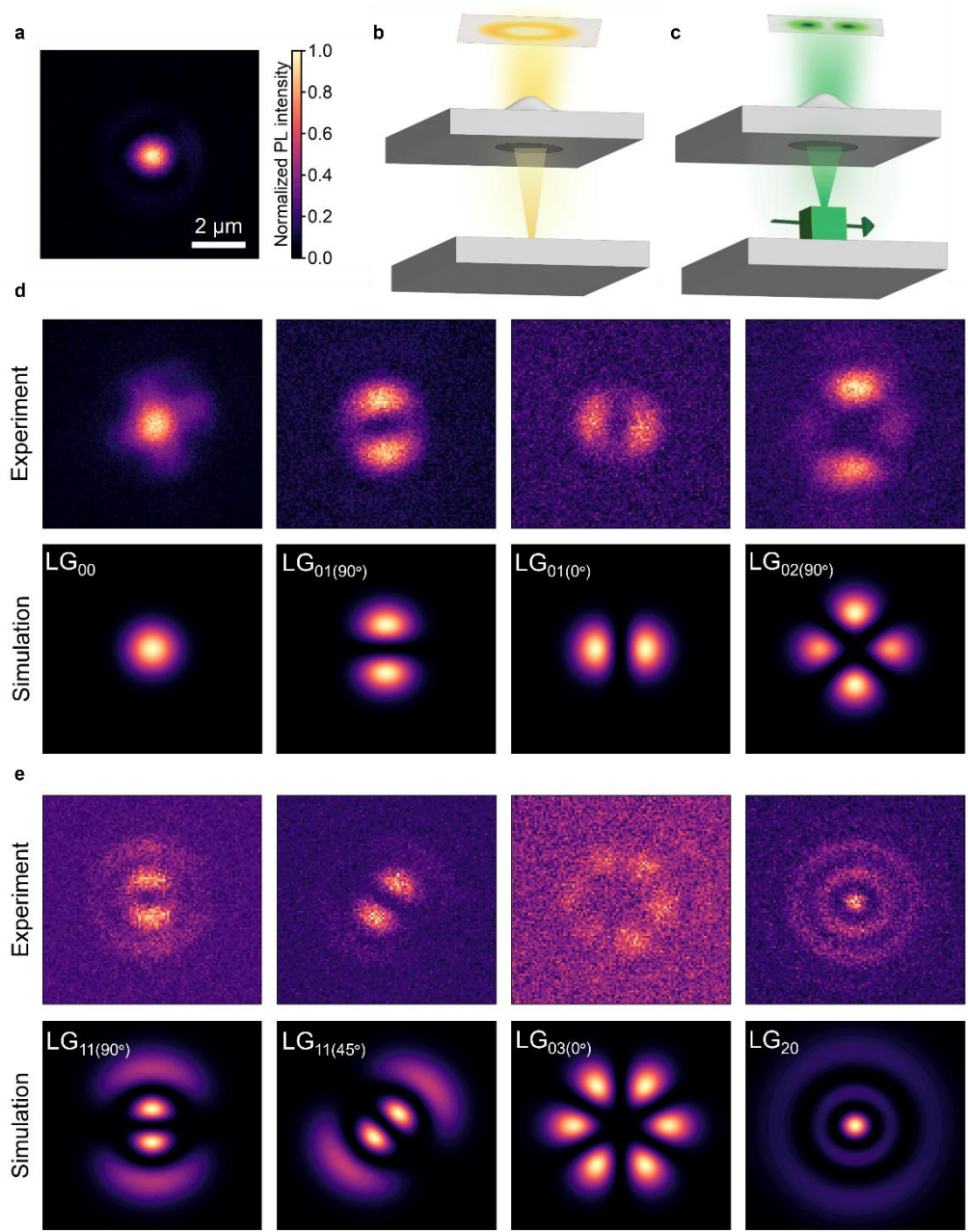

**Fig. 3 | Real-space emission profile of different order Laguerre-Gaussian modes: experiment versus simulation. a,** Real-space emission profile of single QD emission without cavity. **b,** Illustration of the cylindrically symmetric LG modes in the cavity, e.g. a donut shape for the case of $LG_{01}$. **c,** When a linearly polarized emitter is placed in the cavity, the linear polarization couples to both left- and right-helical photon modes, effectively leading to dipole pattern in the emission of $LG_{01}$. **d, e,** Real-space mode profiles (same length scale as (a)) of different order $LG_{nl}$ modes: experiment (top row) and simulation (bottom row). In **(d)**, the emission patterns $LG_{00}$, $LG_{01(90°)}$, $LG_{01(0°)}$ and $LG_{02(90°)}$ are obtained

from a single QD while in-situ tuning the cavity resonance. In **(e)**, $LG_{11(90°)}$ and $LG_{11(45°)}$ were obtained from the same QD (but different QD than (d)) when tuning the cavity, whereas $LG_{03(0°)}$ and $LG_{20}$ come from different QDs each.

| # QD | Temp. (K) | $\tau_{fast,out}$ (ps) | $\tau_{fast,in}$ (ps) | $F_P = \tau_{fast,out}/\tau_{fast,in}$ | $\tau_{slow,out}$ (ps) | $\tau_{slow,in}$ (ps) | $A_{in}/A_{out}$ |
|---|---|---|---|---|---|---|---|
| 1 | 6 | 68 ± 1 | 28.1 ± 0.3 | 2.42 ± 0.04 | 258 ± 7 | 313 ± 7 | 2.8 |
| 2 | 6 | 153 ± 3 | 36.4 ± 0.4 | 4.20 ± 0.09 | 1060 ± 80 | 330 ± 6 | 5.3 |
| 3 | 6 | 111 ± 2 | 30.9 ± 0.4 | 3.59 ± 0.08 | 720 ± 42 | 278 ± 5 | 1.9 |
| 4 | 50 | 484 ± 3 | 39.1 ± 0.4 | 12.4 ± 0.1 | - | 278 ± 1 | 10.0 |
| 5 | 50 | 396 ± 4 | 42.5 ± 0.7 | 9.3 ± 0.2 | - | 278 ± 2 | 14.1 |
| 6 | 50 | 520 ± 3 | 28.7 ± 0.2 | 18.1 ± 0.2 | - | 255 ± 1 | 3.6 |

**Extended Data Table 1 | Summary of the fitted decay times, Purcell factors and intensity ratios for the presented QDs.** The decay traces for QD#1-3 (at 6 K), both inside and outside the cavity, are fitted using a double-exponential function with time constants $\tau_{fast}$ and $\tau_{slow}$. For QD#4-6 (at 50 K), the traces inside the cavity are fitted with a double-exponential function, while those outside the cavity are fitted with a single-exponential function. The offset of the fitting functions is fixed to the average of the counts before the rising time. The Purcell factor, $F_P$, can be obtained from the ratio $\tau_{fast,out} / \tau_{fast,in}$. The intensity ratio $A_{in} / A_{out}$ corresponds to the spectrally integrated emission intensity enhancement inside versus outside the cavity.

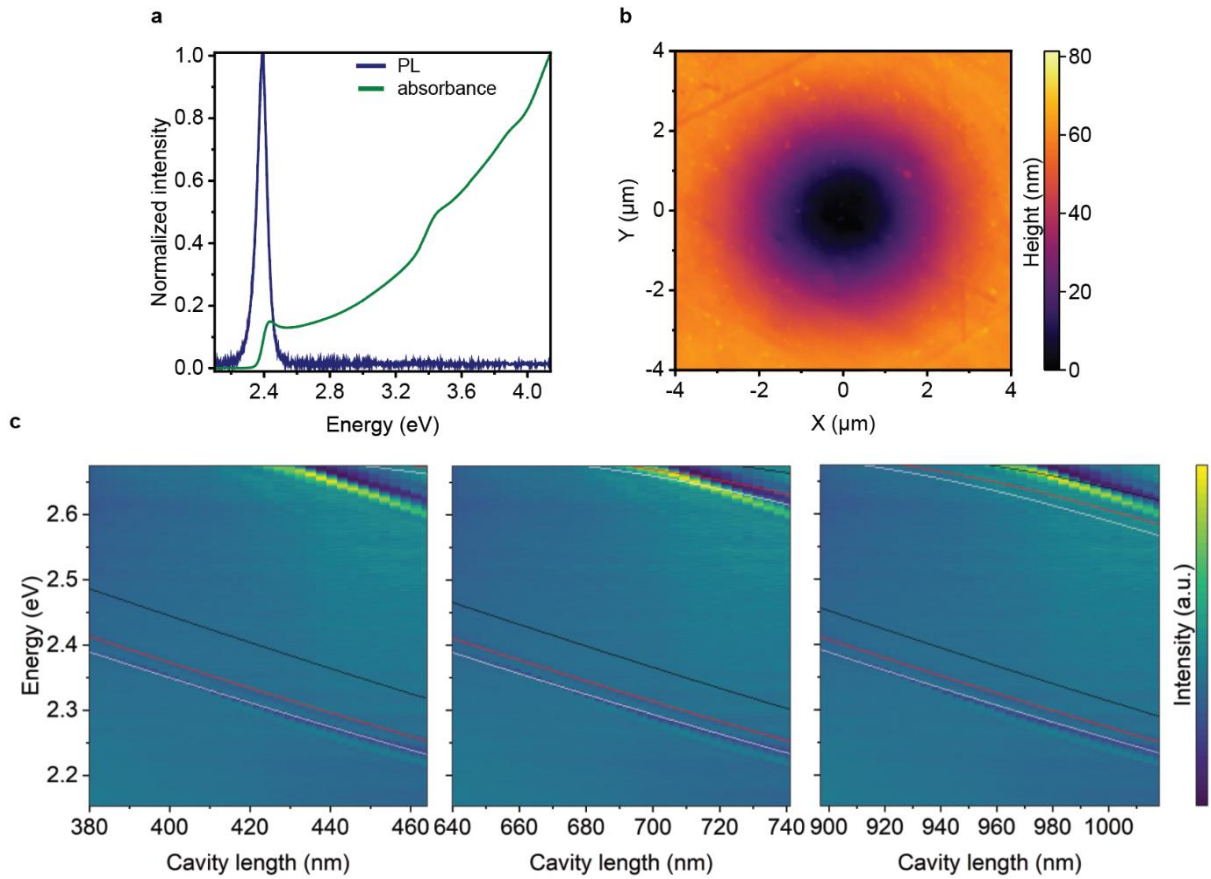

**Extended Data Fig. 1 | Optical properties of CsPbBr$_3$ QDs and cavity characterization. a,** Normalized PL (blue curve) and absorption (green curve) spectrum of 25 nm sized QDs dispersed in solution and measured at room temperature. **b,** Atomic force microscopy image of the Gaussian-shaped deformation, showing a FWHM of ~ 4 μm and a depth of ~ 60 nm. **c,** Comparison between the white-light reflection measurement (colormap) and transfer-matrix simulations for planar modes of different longitudinal order (black lines) to extract the correct range of cavity length. To account for the Gaussian potential, each planar mode of different longitudinal order is horizontally shifted with two different offsets to overlap with both the LG$_{00}$ and LG$_{01}$ modes. These shifted planar modes are displayed as white and red lines, respectively. The shifts for the LG$_{00}$ and LG$_{01}$ are taken from the eigenvalues obtained by solving the 2D Schrödinger equation, assuming a Gaussian potential with a spatial depth of 60 nm and a FWHM of 4 μm. The spatial depth is converted into energy by evaluating the energy shift of a planar mode when the cavity length is varied by 60 nm. Since the slope of a planar mode depends on the longitudinal order and, therefore, on the cavity length, the potential depth varies across

the three panels: 115 meV (left), 94 meV (centre) and 80 meV (right). The resulting eigenvalues are divided by the corresponding potential depth and scaled by 60 nm to retrieve the spatial shifts.

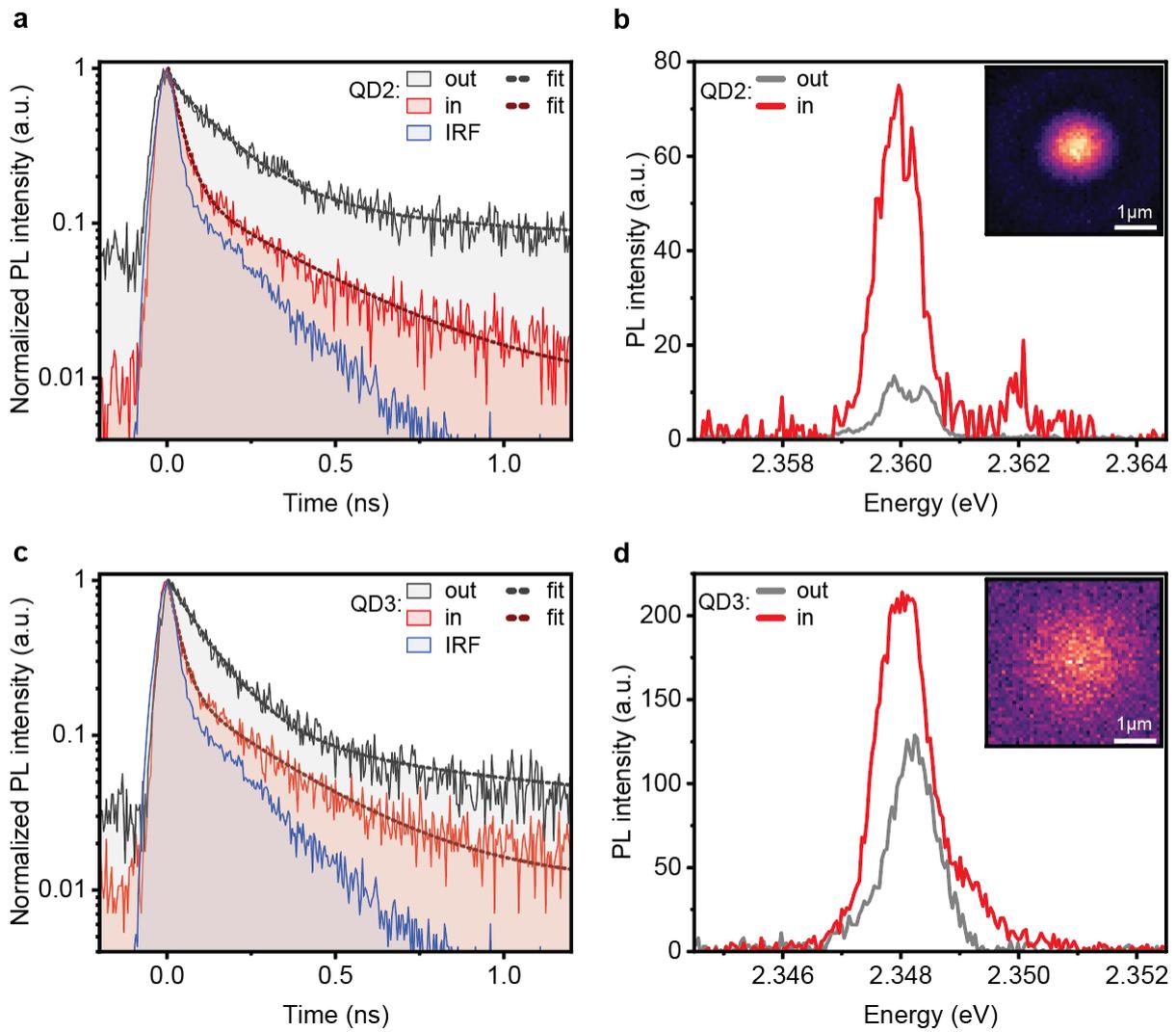

**Extended Data Fig. 2 | Optical study of QD#2, 3 at 6 K. a, b,** Decay traces **(a,c)** and PL spectrum **(b,d)** of QD#2 **(a,b)** and QD#3 **(c,d)** placed inside (red) and outside (black) the cavity, with the IRF (blue) as reference. The insets in **(b,d)** show the measured respective real-space mode profiles.

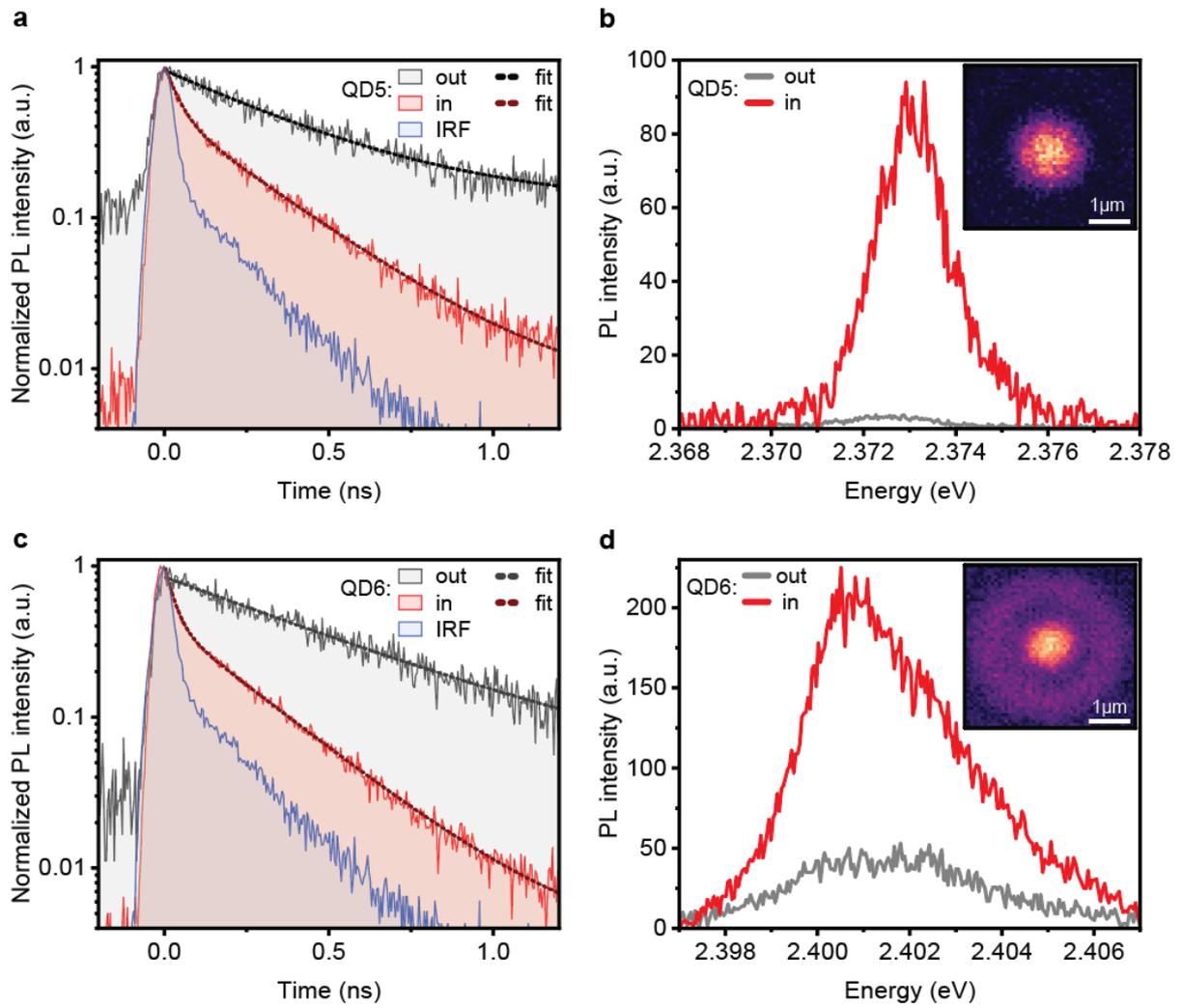

**Extended Data Fig. 3 | Optical study of QD#5, 6 at 50 K. a-d,** Decay traces **(a,c)** and PL spectrum **(b,d)** of QD#5 **(a,b)** and QD#6 **(c,d)** placed inside (red) and outside (black) the cavity, with the IRF (blue) as reference. The insets in **(b,d)** show the measured respective real-space mode profiles.

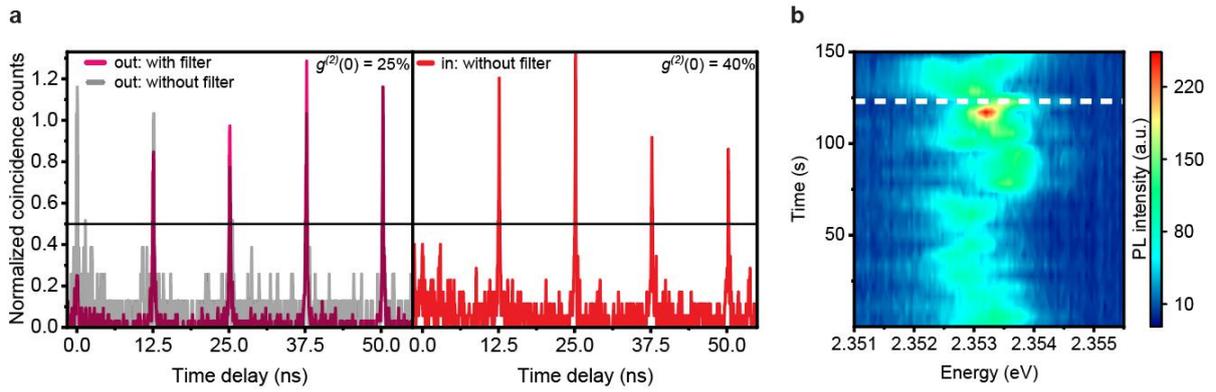

**Extended Data Fig. 4 | $g^{(2)}(0)$ and time-series measurements for QD#1. a,** $g^{(2)}(0)$ measurements outside (left panel) and inside (right panel) the cavity. A $g^{(2)}(0)$ value of 25% is obtained when an additional bandpass filter is used (pink line) while a $g^{(2)}(0)$ close to 100% is measured when the bandpass filter is removed (grey line). Inside the cavity, a $g^{(2)}(0)$ value of 40% is achieved without the use of any filter. These values are extracted from the peak maxima of the raw, uncorrected data. **b,** PL time-series (outside the cavity), acquired with an integration time of 5 s. The dashed white line highlights the time at which the spectrum is shown in Fig. 2b.

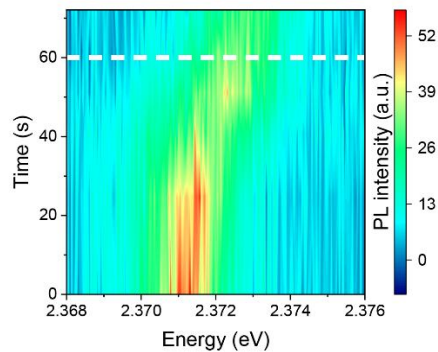

**Extended Data Fig. 5 | Time-series measurements for QD#5.** PL time-series of QD#5. The measurement is acquired when the QD is outside the cavity with an integration time of 10 s. The dashed white line indicates the time at which the spectrum is shown in Extended Data Fig.3b.

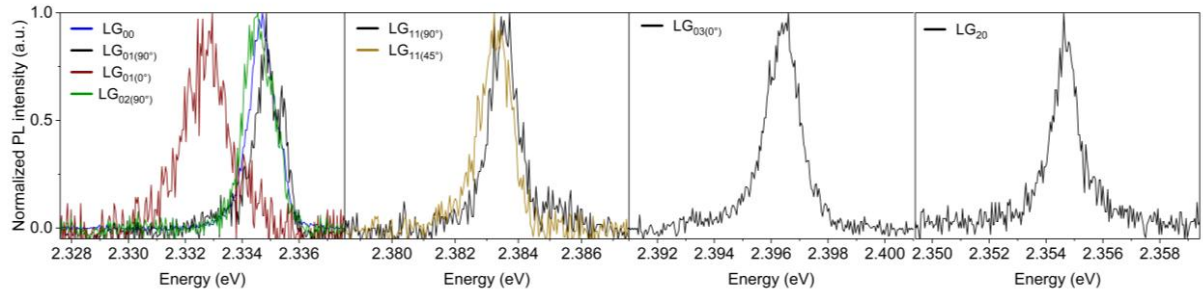

**Extended Data Fig. 6 | Spectra of the LG$_{nl}$ modes in Fig. 3b.** Spectra of the LG$_{nl}$ modes shown in Fig. 3b. The panels are grouped for the same single QDs inside the microcavity, when the LG modes are changed by in-situ tuning.